\documentclass[useAMS,usenatbib]{mn2e}
\usepackage[dvipdfmx]{graphicx}
\usepackage{amsmath,amssymb,natbib,siunitx,booktabs,url}
\usepackage{color}

\newcommand{\lsim}{\lesssim}
\newcommand{\gsim}{\gtrsim}

\newcommand{\lmk}{\left(}
\newcommand{\rmk}{\right)}

\newcommand{\lkk}{\left[}
\newcommand{\rkk}{\right]}
\newcommand{\lla}{\left\langle}

\newcommand{\rra}{\right\rangle}
\newcommand{\so}{M_\odot}
\newcommand{\mch}{{\cal M}}

\newcommand{\bea}{{\begin{eqnarray}}}
\newcommand{\eea}{{\end{eqnarray}}}

\begin{document}

\title[]{ Identifying Disappearance of a White Dwarf Binary with LISA}

\author[]
{Naoki Seto\\
Department of Physics, Kyoto University, Kyoto 606-8502, Japan
}

\date{\today}

\maketitle

\begin{abstract}

We discuss the prospect of identifying a white dwarf binary merger by monitoring disappearance of its nearly monochromatic gravitational wave. For  a ten-year operation of the laser interferometer space antenna (LISA),   the chance probability of observing such an event is roughly estimated to be 20\%.
By simply using  short-term coherent signal integrations, we might  determine the merger time with an  accuracy of $\sim 3$-10\,days.   Also considering its expected  sky localizability  $\sim0.1$-$  0.01 {\rm deg^2}$, LISA  might make an  interesting contribution to the multi-messenger study on a merger event.

\end{abstract}


\section{introduction}

LISA is expected to  detect  $\sim 10^4$ white dwarf binaries (WDBs) in our Galaxy as nearly monochromatic gravitational wave   (GW) sources \citep{1990ApJ...360...75H,2001A&A...365..491N,2012ApJ...758..131N,2018ApJ...854L...1B,2019MNRAS.490.5888L,2022arXiv220306016A}.  Their GW frequencies slightly change  due to the radiation reaction,  and the majority of  them are considered to   merge eventually.  
Given the estimated Galactic merger rate  $R\rm \sim 0.02 yr^{-1}$ (see e.g., \citealt{2001A&A...365..491N}), we might actually have a merger event with a chance probability of  $\sim 0.2 (T/{\rm 10yr})$, during LISA's operation period  $T$.

A WDB merger in our Galaxy will be  an intriguing observational target for multi-messenger astronomy (see e.g., \citealt{2018MNRAS.476.5303S} and also  \citealt{2012ApJ...748...35S,2014MNRAS.438...14D}) and can  help our understanding of white dwarfs (potentially including their explosion process). As an omni-directional GW detector free from interstellar extinction, LISA can provide us  with critical information (e.g. sky locations, orbital phases) for   follow-up electromagnetic (EM) searches (e.g., \citealt{2017MNRAS.470.1894K}).  Therefore, in this paper, we discuss  identification of a WDB merger in LISA's data.   

From various estimations (\citealt{2022arXiv220306016A} and references therein),  the majority of merging WDBs are likely to have total masses less than $1\so$.  In the connection to  type Ia  supernovae, one might be interested in WDBs with the total masses larger than the Chandrasekhar mass. However, their merger rate is estimated to be  $\sim5-7$ times smaller than the total rate quoted above (\citealt{2001A&A...365..491N}, see also \citealt{2018MNRAS.476.2584M}). 

According to recent numerical simulations with a careful treatment for the initial conditions (\citealt{2011ApJ...737...89D} see also \citealt{1990ApJ...348..647B,1992ApJ...401..226R,2006ApJ...643..381D,2009A&A...500.1193L,2012ApJ...746...62R,2013ApJ...770L...8P}),  a typical merging WDB   will continue to emit nearly monochromatic GW,  until just before the donor star is tidally disrupted and the system virtually turns off GW emission.
On the basis of this overall GW emission pattern, we examine a forceful  approach for the identification of a WDB merger by repeatedly checking the resultant disappearance of its nearly monochromatic GW. 
We pay an attention to the estimation of the merger time, which will be important for EM observations.

This paper is organized as follows. In \S 2, we briefly describe the evolution of a merging WDB and the associated GW emission. 
In \S 3, we discuss LISA's observation for the GW  signal, including estimation of the merger time with matched filtering analysis. In \S 4, we discuss related topics such as the cases with other space interferometers.  \S 5 is devoted to a short summary. 
\section{Evolution of a merging WDB}

\subsection{Nearly Monochromatic Waves}
As an approximation to a detached WDB emitting a nearly monochromatic GW,  we briefly discuss a circular binary composed by two point masses $m_1$ and $m_2$ ($m_1\le m_2$).  
Using the Kepler's law, we can write down the wave  frequency 
\begin{eqnarray}
f=\pi^{-1}G^{1/2}(m_1+m_2)^{1/2}a^{-3/2} \label{kf}
\end{eqnarray}
with the orbital separation $a$ and the gravitational constant $G$. At the quadrupole order, the angular averaged strain amplitude is written as 
\begin{eqnarray}
h=\frac{8G^{5/3}\mch^{5/3}\pi^{2/3}f_{\rm }^{2/3}}{5^{1/2}c^4 d}\label{h}
\end{eqnarray}
with the distance  $d$  to  the binary (below fixed at the representative distance  $d=10$kpc) and the speed of light $c$ \citep{2019CQGra..36j5011R}.
Here we put the chirp mass 
$\mch\equiv (m_1m_2)^{3/5}(m_1+m_2)^{-1/5}$.

The frequency evolution due to the radiation reaction is written as 
\begin{eqnarray}
\frac{df}{dt}&=&\frac{96G^{5/3}\pi^{8/3}\mch^{5/3} f^{11/3}}{5c^5}\label{df}\\
&=&2.5 \times 10^{-6}  \lmk \frac{f}{\rm 20.8mHz} \rmk^{11/3}   \lmk \frac{\mch}{0.38\so} \rmk^{5/3} {\rm Hz\,yr^{-1}}. \nonumber
\end{eqnarray}
The tidal interaction between the two star  will modify our expressions (\ref{kf})-(\ref{df}) to some extent (e.g., $\sim 10\%$ for  Eq. (\ref{df}) as in \citealt{2021MNRAS.500L..52W}).  However, in this paper, we are mainly interested in the identification of a WDB merger event from LISA's data, by just using  the disappearance of its GW (not  trying to accurately estimate  e.g., its chirp mass).  Later, we use Eq. (\ref{h})  in  this context, and it will work reasonably well, up to very close to the final merger \citep{2011ApJ...737...89D}. 

Given the merger rate (or more appropriately the merger flux) $R$ of Galactic WDBs, we can apply the continuity equation in the frequency space and approximately evaluate   their  frequency distribution $dN/df$  as
\begin{eqnarray}
\frac{dN}{df}&\sim& R  (df/dt)^{-1}\\
&\sim& 3.4\times 10^6 \lmk \frac{R}{\rm 0.02 yr^{-1}} \rmk\lmk \frac{f}{\rm 4mHz} \rmk^{-11/3} {\rm Hz^{-1}}. \label{dn2}
\end{eqnarray}
This expression will be valid  in the frequency  regime $f\lsim  10$mHz where the merger flux $R$ is expected to be nearly constant \citep{2022PhRvL.128d1101S}. 
Here, for simplicity, we ignored the chirp mass distribution  and  put $\mch=0.38\so$.  From an integral of  Eq. (\ref{dn2}),  the number of Galactic WDBs above $\sim  4$mHz is roughly estimated to be $N(>{\rm 4mHz})\sim 5\times 10^3$.

\subsection{Transition around the Merger}

Due to the radiation reaction, the orbital separation of a detached WDB  decreases gradually and the less massive donor $m_1$  will eventually fill its Roche-lobe, initiating mass transfer  to the accreter $m_2$ \citep{1967AcA....17..287P,2004MNRAS.350..113M}.  The time evolution during this mass transfer phase  has been examined by numerical simulations \citep{2011ApJ...737...89D}. 
A typical merging  WDB (like those analyzed below) will continue to emit nearly monochromatic GW  (roughly described in the previous subsection), until the donor loses a small fraction of its mass due to an unstable mass transfer (see e.g., Figs 7 and 17 in Dan et al. 2011).   

Then, in a short transitional period $\Delta t_{\rm t}$ (corresponding to $\lsim 10$ wave cycles), the donor is tidally disrupted, and  the system will soon settle down into a nearly axisymmetric configuration, virtually stopping GW emission (\citealt{2011ApJ...737...89D}, see also  \citealt{2021ApJ...906...29Y,2023arXiv230305519M}).
We define the merger time $t_{\rm m}$ as the beginning of the transition period $\Delta t_{\rm t}$.   One of our primary objectives below is to clarify how well we can observationally determine the merger time $t_{\rm m}$ only  with LISA.  

We also define  the merger frequency $f_{\rm m}$ as the GW frequency  at end of the nearly monochromatic emission (just before the transition period). Similarly, we put the associated characteristic strain amplitude $h_{\rm m}$  which is obtained by plugging in  the final frequency $f=f_{\rm m}$ in Eq. (\ref{h}).

Next, following \citet{{2004MNRAS.350..113M}}, let us approximately  estimate the merger  frequency $f_{\rm m}$, by using the condition of 
the Roche-lobe filling of the  less massive (donor) star $m_1$. 
We apply the mass radius relation $r(m_1)$ for completely degenerate helium   given in   Verbunt \&  Rappaport (1988), ignoring (potentially existing) diffuse outer envelope.  Meanwhile the Roche-lobe radius is approximately given by \citet{1967AcA....17..287P} as 
\begin{eqnarray}
R_L=2\cdot 3^{-4/3} m_1^{1/3} (m_1+m_2)^{-1/3}a \label{rl}.
\end{eqnarray}
Then  we put  $R_L=r(m_1)$ for the donor $m_1$ at the onset of the mass overflow.  From Eqs. (\ref{kf}) and (\ref{rl}), the merger frequency is written as 
\begin{eqnarray}
f_{\rm m}(m_1)=\frac{2^{3/2}G^{1/2}m_1^{1/2}}{9\pi r(m_1)^{3/2}}, \label{fm}
\end{eqnarray}
depending only on the mass  $m_1$  of the donor.

In this paper, considering the mass distribution of WDBs, we examine the three WDB models presented in Table 1. 
We set the model A as the base model for our study with  the lower-mass comparative  model B. The reference model C  has the total mass  of $1.5\so$.  
In Table 1, we present their merger frequencies (\ref{fm}) . From  a comparison with a more detailed numerical study in Dan et al. (2011), our rough estimation is  expected to have accuracy within $\sim 15\%$   for the presented donor masses $m_1$.   

At the frequency $f_{\rm m}=20.8$mHz,  under the point particle approximation  the base model A has the radiation reaction time scale $f/{\dot f}\sim 10^4$yr. Therefore, even $\sim 10$yr before the merger, we approximately have its frequency and amplitude as  $f\sim f_{\rm m}$ and  $h\sim h_{\rm m}$.  

\begin{table}
\caption{Nearly monochromatic GWs from  merging WDBs (A: our base model,  B: the lower mass model, C: the higher mass model).  We put the donor mass $m_1$ and the accreter mass $m_2$.  The binary merges around the frequency $f_{\rm m}$.   We show the signal-to-noise ratios $\rho$ during the final two years before the merger with LISA (at the distance of $d=10$kpc).  We present the short-term  integration period $T_{\rm s}$ to get signal-to-noise ratio  of $\rho_{\rm th}=7.5$.  The corresponding rotation cycles $f_{\rm m}\cdot T_{\rm s}$ are also presented.  }
 \centering
 
\begin{tabular}{@{}l|lll@{}}
\toprule
  model &   A  (base)   & B   & C  \\ \midrule
 $m_1$ & 0.4 $\so$  &  0.3 $\so $ &   { 0.6} $\so $       \\
 
$m_2/m_1$ & 1.2  & 1.2  &   1.5       \\

 \midrule

$f_{\rm m}$  & 20.8\,mHz  & 15.0\,mHz   &   35.1\,mHz        \\

$\rho$ ($T=$2yr)   & $85 $ & 47 & $210 $ \\

$T_{\rm s}$ ($\rho_{\rm  th}=$7.5)   &  5.7\,day & 18.6\,day & 0.93\,day \\

$f_{\rm m}\cdot T_{\rm  s}$    & 10000 & 24000 & 2800\\
 \bottomrule
\end{tabular}
\end{table}


\subsection{Evolutionary Stages}

On the basis of the previous two subsections, we next divide the time evolution  of a merging WDB into the four stages (i)-(iv), as a preparation for  the matched  filtering analysis mainly discussed in the next section.

\begin{figure}
\vspace{-.2cm}
 \includegraphics[width=1.22\linewidth]{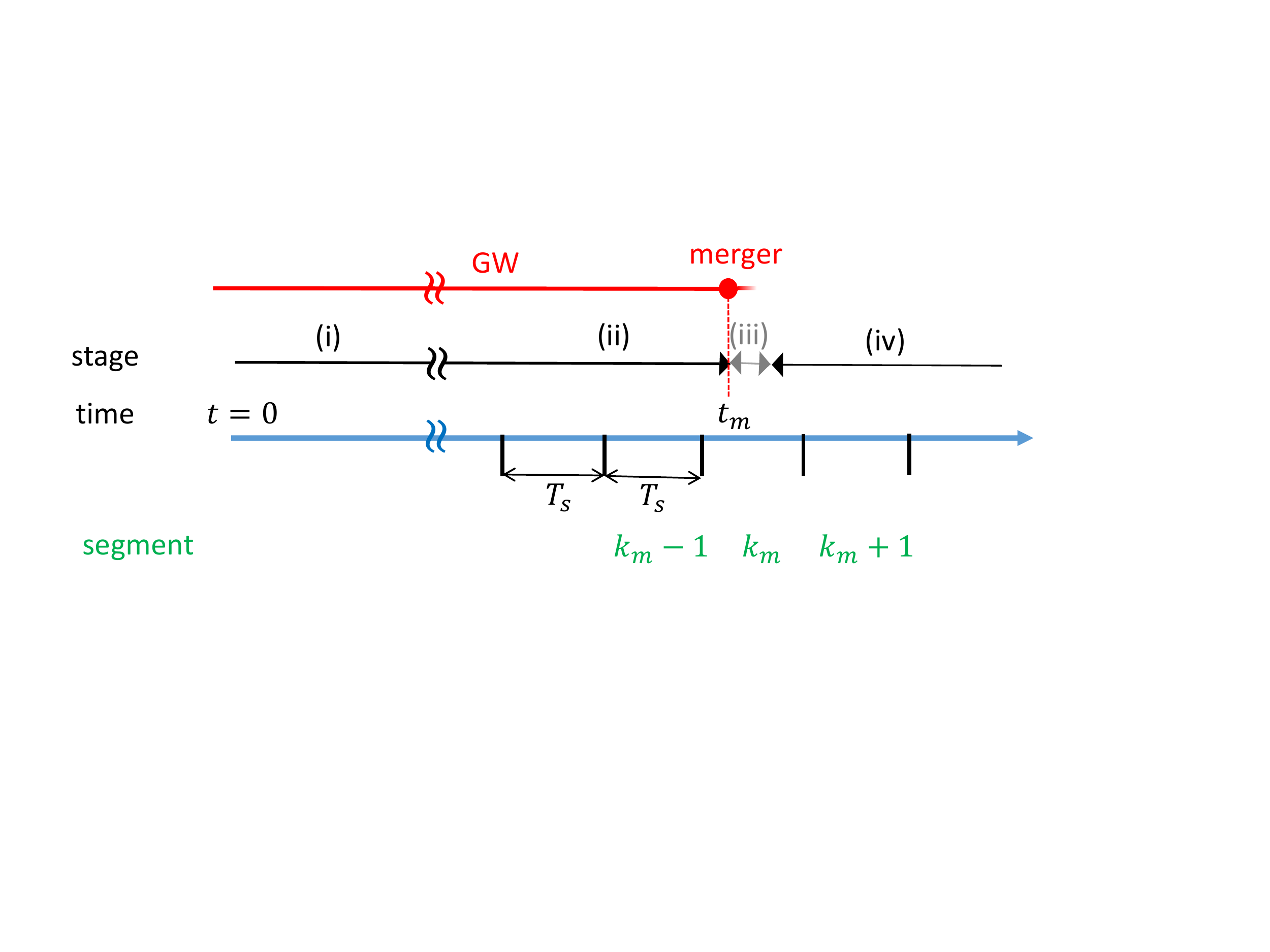} 
 \vspace{-2.6cm}
 \caption{Schematic diagram for observation of a GW signal from a  merging WDB.  We set the time origin $t=0$ at the beginning of LISA's operation and put the merger time  by $t=t_{\rm m}$.  We divide the time sequence into the four stages (i)-(iv) with the duration $\Delta t_{\rm p}$ and $\Delta t_{\rm t}$ respectively for (ii) and (iii).    For the identification of the merger time $t_{\rm m}$, we introduce the time segmentation with an appropriately selected  duration $T_{\rm s}$.   The merger  time $t_{\rm m}$ is in the segment $k=k_{\rm m}$.  
 }  \label{fig:volume}
\end{figure}

For coherent signal integration with the matched filtering analysis, the phase of the target GW should be accurately characterized by a limited number of fitting parameters (e.g., the polynomial expansion with the derivative coefficients   $f^{(n)}$ at some epoch). Unlike a well separated WDBs,  even in  the time region of nearly monochromatic GW emission,  a WDB close to the merger might have  complicated GW phase modulations which should be handled with  a short-term phenomenological phase modeling.  As a precaution for such a potential difficulty, we conservatively divide the time region of nearly monochromatic emission into two stages (i) and (ii).  We put $\Delta t_{\rm p}$ as the total duration of the late monochromatic stage (ii).  
Sorting out the temporal sequence, around the merger time $t_{\rm m}$,  we now have   the following four stages.

(i) Early monochromatic stage at $t<t_{\rm m}-\Delta t_{\rm p}$.  The binary emits nearly monochromatic GW at the frequency $\sim f_{\rm m}$ with the amplitude $\sim h_{\rm m}$. Its intrinsic GW phase can be coherently described by a simple expression.

(ii) Late monochromatic stage at $t_{\rm m} -\Delta t_{\rm p}<t<t_{\rm m}$. The binary continues to emit nearly monochromatic wave at $\sim f_{\rm m}$ and $\sim h_{\rm m}$. We need phenomenological phase model with  a short integration period. 

(iii)  Transition stage at $t_{\rm m} <t<t_{\rm m}+\Delta t_{\rm t}$.  The disruption of the donor proceeds rapidly  and the emitted  GW no longer has regular monochromatic pattern,    due to  hydrodymanical effects.

(iv) Post-merger stage at $t_{\rm m}+\Delta t_{\rm t}<t$.  Virtually no GW is emitted, as the merged system settles down to a nearly stationary and axisymmetric configuration.

Here we can practically assume  $\Delta t_{\rm p}\gg \Delta t_{\rm t}$ (otherwise we do no need to introduce the late stage (ii)).
In contrast to the transition duration $\Delta t_{\rm t}$ deduced from existing numerical simulations,   the magnitude of $\Delta t_{\rm p}$ is currently unclear.   We need to  accurately follow  the  longer-term dynamical evolution at high mass resolution.  Numerical  approach  would be highly demanding, and analytical study might be more advantageous.  
Observational studies on known short-period WDBs might be also useful (see e.g. \citealt{2021ApJ...912L...8S,2023MNRAS.518.5123M}  for RX J0806.3+1527).
 Below, we leave the duration $\Delta t_{\rm p}$ as a free parameter, just assuming $\Delta t_{\rm p}\ll 1$yr.  
We should also comment  that the boundary between (i) and (ii)  is not mathematically distinct.  It is rather a practical one and depends  on the complexity of the phase modeling allowed for the stage (i). 

\section{Observation with LISA}

Next we discuss matched filtering analysis with  LISA. We set the origin of the time coordinate $t=0$ at the beginning of LISA's operation.  Given its planned operation period 4-10 yr, we below set $t_{\rm m}=2$yr as a conservative average.  For  the purpose of concise illustration,  we are assumed  to make two kinds of data analysis:  the long-term search and the  segmented short-term search. 

The purpose of the long-term search is to confidently detect a WDB and determine its basic parameters, using the data at the early monochromatic stage (i).  This search would be similar to the signal analysis for a standard (non-merging) WDB.   Using the results of the long-term search, we subsequently perform the segmented  short-term searches  for identifying the disappearance of its nearly monochromatic GW at the merger.  In reality, we can chronologically mix these two searches.

\subsection{The Long-Term Search}
Throughout the early monochromatic stage (i), the GW signal would be coherently searched with manageable  numerical costs.  
 The total signal-to-noise ratio $\rho$ can be estimate as
\begin{eqnarray}
\rho=\frac{h_{\rm m}\sqrt{T}}{S_n(f_{\rm m})^{1/2}} \label{rho}
\end{eqnarray}
for the integration  time $T$. Here, for simplicity,  we ignore the angular dependence of the GW strain \citep{1998PhRvD..57.7089C}. It is straightforward to  include it in our analysis below. 

As presented in  Table 1 for $T=t_{\rm m}=2$yr,  the expected signal-to-noise ratio in the stage (i) is $\rho=O(100)$ and much larger than those for typical Galactic  LISA sources existing  at lower frequencies.   
For the base model A (up  to the end of the stage (ii)),  the signal-to-noise ratio squared $\rho^2$ increases unity in every $\sim 180$ wave cycles, corresponding to $\sim$0.1 day ($\sim430$ cycles for B and $\sim 50$ cycles for C). 
Therefore, even though the catastrophic stage (iii) is interesting,  this stage is  likely to  have a negligible contribution to the signal-to-noise ratio, given its rotation cycles of $\lsim 10$ \citep{2011ApJ...737...89D}.   For our signal analysis scheme, we can omit the stage (iii) and effectively regard that the nearly monochromatic stage (ii) is directly followed by the quiet stage (iv) at the merger  time $t_{\rm m}$.

Using the high quality signal in the stage (i), we can securely  confirm the existence of the binary. We can also  estimate  its nearly monochromatic frequency $f_{\rm m}$ and other basic parameters including the amplitude  $h_{\rm m}$.  For instance, the sky position of the binary can be determined with the aid of the Doppler modulation \citep{1998PhRvD..57.7089C}. For $T\gsim 2$yr,  we can apply an  asymptotic scaling relation for the area of the error ellipsoid \citep{2002ApJ...575.1030T}
\begin{eqnarray}
\Delta \Omega\sim0.01  \lmk \frac{\rho}{85} \rmk^{-2}\lmk \frac{f}{\rm 20.8mHz} \rmk^{-2} {\rm deg^2}. 
\end{eqnarray}     
The area becomes  $\rm \sim 0.06 deg^2$ for the model B.
The high sky localization would be beneficial for  associated EM observations.

\subsection{Segmented Short-Term Search}

Let us suppose that we have already done the  long-term search for a WDB, in the middle of its  early monochromatic stage (i).  Now, we divide the upcoming data streams into segments of a duration $T_{\rm s}$ as shown in Fig. 1.   We  assign the label $k$ and put $\rho_k$ for the corresponding  signal-to-noise  ratio obtained after the short-term coherent matched filtering.  The total number of the segments increases, as  LISA's operation  period becomes longer.  

Our immediate goal is to observationally determine the specific segment $k_m$ which contains the merger time $t_{\rm m}$ (see  Fig. 1). To this end, we  appropriately set the threshold $\rho_{\rm th}$ and the duration $T_{\rm s}$, so that we  can reliably have $\rho_k\ge \rho_{\rm th}$ for earlier segments $k<k_m$  and $\rho_k< \rho_{\rm th}$ for  later segments  $k>k_m$, due to the existence of the nearly monochromatic wave before the merger time $t_{\rm m}$.

Here,  in  terms of statistical tests,  we want to suppress  the false dismissal rate $P_{\rm FDR}$ for $k<k_m$ and  the false alarm rate $P_{\rm FAR}$ for $k>k_m$, both caused by the flukes of the noise realizations \citep{2003PhRvD..67b4016B}. We should notice that, for the target WDB,  presupposing the high-quality information from the already analyzed data (in the stage (i)), the implication of these  two rates  are somewhat different from the blind short-lived binary search with the current LVK network \citep{2021arXiv211103606T}.

For a segment at $k>k_m$, assuming Gaussian noise, the false alarm rate is estimated to be
\begin{eqnarray}
P_{\rm FAR}= \frac{\cal N}{\sqrt{2\pi}} \int_{\rho_{\rm th}}^\infty  {\exp\lkk -\frac{\rho^2}2\rkk } d\rho=\frac{\cal N}2{{\rm erfc}\lkk \frac{\rho_{\rm  th}}{2^{1/2}}\rkk}
\end{eqnarray}
with the effective number of the templates  $\cal N$ and the complementary error function $\rm erfc(\cdot)$.
Note that the function ${{\rm erfc}\lkk {\rho_{\rm  th}}/{2^{1/2}}\rkk}/2$ depends very strongly on $\rho_{\rm th}$ and we have  $10^{-2},10^{-3}, 10^{-5}$ and $10^{-7}$ respectively for 
$\rho_{\rm th}=2.3$, 3.1, 4.3 and 5.2.
The effective number of templates $\cal N$ depends on the  possible phase models (and the integration time $T_{\rm s}$). After all, in our pilot study,  we shifted the uncertainty of the stage (ii) to that of the template number $\cal N$.   In the actual data  analysis, we  can use the information of the earlier segments, for empirically  extending the phase models. At present, on the basis of the strong dependence on $\rho_{\rm th}$, we adopt the reference value $\rho_{\rm th}=5$ for $P_{\rm FAR}=10^{-2}$.   If we actually have $\Delta t_{\rm p}\ll T_{\rm s}$, our setting $\rho_{\rm th}=5$ will result in a very conservative choice.

Next we discuss the false dismissal rate for an earlier segment $k<k_m$. Using the results (e.g., $h_{\rm m}$ and $f_{\rm m}$) from the preceding long-term search, we can estimate  the expectation value 
\begin{eqnarray}
\lla \rho_k\rra =\frac{h_{\rm m}\sqrt{T_{\rm s}}}{S_n(f_{\rm m})^{1/2}} \label{eb}
\end{eqnarray}
as a function of the segment duration $T_{\rm s}$.  Then, by adjusting the duration  $T_{\rm s}$,  we can evaluate the false dismissal rate as
 \begin{eqnarray}
P_{\rm FDR}&=& \frac{1}{\sqrt{2\pi}} \int^{\rho_{\rm th}}_{-\infty}  {\exp\lkk -\frac{(\rho-\lla \rho_k\rra )^2}2\rkk } d\rho\\
&=&\frac{1}2{{\rm erfc}\lkk \frac{\lla \rho_k\rra -\rho_{\rm  th}}{2^{1/2}}\rkk}.
\end{eqnarray}

For $P_{\rm FDR}\sim 10^{-2}$, we have 
\begin{eqnarray}
\lla \rho_k\rra \simeq \rho_{\rm th}+2.5=7.5.
\end{eqnarray}
We can inversely  obtain the segment duration $T_{\rm s}$ by 
\begin{eqnarray}
T_{\rm  s}= \lla\rho_{k}\rra^2 S_n(f_{\rm m}) h_{\rm m}^{-2}. \label{tth}
\end{eqnarray}

In Table 1, we present the numerical results for $T_{\rm s}$, together  with the corresponding rotation cycles $f_{\rm m}\cdot T_{\rm s}$. For the base model, we have $T_{\rm s}=5.7$ day and 10000 cycles. 

From the arguments so far, we can easily  see that, for the critical segement $k=k_m$,  we have  $\rho_k\ge \rho_{\rm th}$ or $\rho_k< \rho_{\rm th}$, depending on the location of the merger time   $t_{\rm m}$ in the segment.

\subsection{Estimation of the Merger Time}
Next we discuss the estimation of the merger time $t_{\rm m}$ from the growing list $\{\rho_k\}$ obtained after repeating the short-term matched filtering. To proceed in a dualistic manner, we map the signal-to-noise ratios $\rho_k$ by
\begin{eqnarray}
F_k\equiv \theta(\rho_k-\rho_{\rm th}) \label{map}
\end{eqnarray}
with the step function $\theta(\cdot)$.  We will have $F_k=1$ for $k<k_m$ and $F_k=0$ for $k>k_m$ ($F_k=1$ or 0 for $k=k_m$).  Then,  the list $\{F_k\}$ should have  the following form 
\begin{eqnarray}
\{F_k\}=\{ 1,1\cdots,1,\underline{1,0}, 0, 0, 0,  \cdots \}.
\end{eqnarray}
The merger time $t_{\rm m}$ should contained  in the two specific segments marked with the underline.   Therefore, the  half width $\delta t_{\rm m}$ of the estimation error becomes 
\begin{eqnarray}
\delta t_{\rm m}=2^{-1}(2 T_{\rm s})=T_{\rm s}.
\end{eqnarray}

So far, we have fix the segmentation. By relaxing this setting, we can further analyze the two candidate segments above.  More specifically, we put a slidable segment characterized by the starting  time $x$ (keeping the duration at $T_{\rm s}$) as
\begin{eqnarray}
k(x)\equiv [x,x+T_{\rm s}].
\end{eqnarray}
We define the associated signal-to-noise ratio $\rho(x)$ for this segment.  We can identify the maximum value $x=x_{\max}$ which satisfies $F[\rho{(x)}]=1$.  Similarly, we read the minimum value $x=x_{\min}$   for $F([\rho{(x)}]=0$ with $x_{\max}\le x_{\min}$.  Here, the inequality can be caused by an intricate pattern of the function $F[\rho{(x)}]$  due the noise.  The merger time $t_{\rm m}$ should be contained  simultaneously  in the two segments  $k(x_{\max}) $ and $k(x_{\min})$. Taking their overlapped part,  we can estimate the merger time with a half width of 
\begin{eqnarray}
\delta t_{\rm m}\le T_{\rm s}/2,
\end{eqnarray}
corresponding to $\lesssim 3$ days  for our base model. 

The estimated magnitude  $\delta t_{\rm m}$ is much larger  than the transition period  $\Delta t_{\rm t}$ around the merger.  Therefore, as mentioned earlier, we  practically do not need to too rigidly define the merger time $t_{\rm m}$ (e.g., before or after the transition stage (iii)).

\section{Discussion}

So far, we have discussed matched filtering analysis for a merging Galactic WDB with LISA. Here, from a broader perspective, we mention related topics and potential  extension of this work.

\subsection{GW data analysis}
We have assumed $\Delta t_{\rm p}\ll $1yr, for the duration  $\Delta t_{\rm p}$ of the late monochromatic stage (ii).  Currently, it is not straightforward to solidly estimate the duration $\Delta t_{\rm p}$ introduced for precautionary purpose.   As already commented, we can rather evaluate the potential phase evolution, by empirically using the existing earlier data of the binary (smoothly from the stage (i)).  Similarly, the number of  effective  templates $\cal N$  will be estimated reasonably well at the actual data analysis in the stage (ii).

In a very pessimistic  scenario with $\Delta t_{\rm p}\gsim $2yr, we  might not have the simple stage (i), which has played a key role  in our data analysis procedure.   A semi-coherent search could be a powerful option, as currently applied  to unknown pulsar search with the LVK network \citep{2022PhRvD.106j2008A}. 
GW from a WDB could be observed at much higher signal-to-noise ratio with much smaller rotation cycles. Therefore, depending on the phase irregularity, our task could be much easier than the unknown pulsar search with the LVK network.

At estimating the merger time $t_{\rm m}$, we have  used a matched filtering analysis just for checking the disappearance of a  nearly monochromatic GW signal (without decoding its detailed structure).
In  reality, the wave phase might be relatively simple  with  a precursor signature for the forthcoming merger. Then, we might predict the merger time $t_{\rm m}$ by carefully performing matched filtering analysis. 
If  this is the case, our estimation $\delta t_{\rm m}$ should be regarded as a very conservative bound.  On another front,  while we have introduced the  mapping (\ref{map}) for our simple dualistic argument,  the original function $\rho(x)$ can be directly analyzed for the estimation of the merger time $t_{\rm m}$.

{Note that, if the operation period of LISA is longer than $\sim 1$yr,  the Galactic confusion noise is negligible in the frequency regime $f\gsim 5$mHz \citep{{2019CQGra..36j5011R,2020PhRvD.101l3021L}}.  Therefore, for a donor mass $m_1\gtrsim 0.3\so$ shown in Table 1, the interference with other binary signals is likely to be insignificant.  However, we have $f_{\rm m}\sim 5$mHz, for a  smaller donor mass  $m_2\sim 0.1\so$.   For such a binary,  we might need to securely suppress the  interference effects at the LISA multi source-analysis.  Consequently, the required segment duration $T_{\rm  s}$ could be longer than that estimated from Eq. (\ref{eb}).}  

 We can try other efficient methods (e.g., the chi-square analysis), particularly in the context of  the low latency analysis. The non-Gaussianity of detector noises might be also worth studying. 

\subsection{Other Space Interferometers}
Here we briefly  comment on  space interferometers other than LISA. 
Since a WDB mainly emits a nearly monochromatic GW, our results scale simply with the instrumental noise spectrum $S_n(f)$, as shown in Eqs. (\ref{rho}) and (\ref{tth}).   

Around 20mHz,  the sensitivities of  Taiji \citep{2018arXiv180709495R} and TianQin \citep{2016CQGra..33c5010L} are quite similar and approximately two  times better than that of  LISA.   Meanwhile, at $\sim$20mHz,  B-DECIGO \citep{2021PTEP.2021eA105K} is designed   to have a sensitivity similar to LISA, but its optimal band is in the higher frequency regime  around 100mHz \citep{2018PTEP.2018g3E01I}.  B-DECIGO and its follow-on mission DECIGO \citep{2001PhRvL..87v1103S} might thereby enable us to detect extra-Galactic merging  WDBs with total masses higher than the Chandrasekhar mass  (possibly relevant for SNIa) \citep{2020A&A...635A.120M,2022ApJ...938...52K}.

\subsection{EM Observation}
When we try to identify  an  EM counterpart to a WDB,   its binary parameters (e.g., those related to orbital phase, sky location and inclination) will be critically useful.  Once  identified, we can continue to  frequently observe the binary with EM telescopes, even after the operation period of LISA. Moreover, we might monitor the transitional merger stage (iii) with the telescopes. Such an  observation will provide us with valuable information on white dwarfs and, possibly,  its explosion mechanism.  The archived data much before LISA's launch might be also useful for studying a WDB in a long time span \citep{2022arXiv221214887D}.

\section{summary}

The merger rate of Galactic WDBs is roughly estimated to be $\rm \sim 0.02 yr^{-1}$. 
In this paper, we discuss identification of a merger event in the data streams of  LISA.   
Until very close its final merger,  a typical WDB will emit nearly monochromatic GW around 20mHz. After a short transition period, it will virtually stop emitting GW.  LISA has potential to detect its nearly monochromatic GW with a signal-to-noise ratio of $O(10^2)$ and determine its sky location within $\sim 0.01$-$0.1{\rm deg^2}$. 
By repeatedly checking the disappearance  of its nearly monochromatic GW, we could robustly estimate its merger time with accuracy of  $\sim3$-10\,days. 
\section*{Acknowledgements}

This work is supported by JSPS Kakenhi Grant-in-Aid for  Scientific Research (Nos.~ 17H06358, 19K03870 and 23K03385).




\bibliographystyle{mn2e}

\end{document}